\documentclass[draftcls,onecolumn,12pt]{IEEEtran}

\usepackage{relsize}
\usepackage{color}
\usepackage{mathrsfs}
\usepackage{eucal}
\usepackage{graphicx}
\usepackage{amssymb}
\usepackage{algorithm}
\usepackage{algorithmic}
\usepackage{eqnarray}
\usepackage{booktabs}
\usepackage{amsbsy}
\usepackage{xparse}

\ifCLASSINFOpdf
\else
\fi
\begin{document}

%
\title{Learning-Aided Deep Path Prediction for Sphere Decoding in Large MIMO Systems}
%
%
%

\author{{Doyeon Weon} and
        {Kyungchun Lee,~\IEEEmembership{Senior Member,~IEEE}}
\thanks{{The authors are with the Department of Electrical and Information Engineering and the Research Center for Electrical and Information Technology, Seoul National University of Science and Technology,  Seoul 01811, Republic of Korea (e-mail: weondy123, kclee@seoultech.ac.kr)}.}
}

\maketitle

\begin{abstract}
In this paper, we propose a novel learning-aided sphere decoding (SD) scheme for large multiple-input--multiple-output systems, namely, deep path prediction-based sphere decoding (DPP-SD). In this scheme, we employ a neural network (NN) to predict the minimum metrics of the ``deep'' paths in sub-trees before commencing the tree search in SD. To reduce the complexity of the NN, we employ the input vector with a reduced dimension rather than using the original received signals and full channel matrix. The outputs of the NN, i.e., the predicted minimum path metrics, are exploited to determine the search order between the sub-trees, as well as to optimize the initial search radius, which may reduce the computational complexity of SD. For further complexity reduction, an early termination scheme based on the predicted minimum path metrics is also proposed. Our simulation results show that the proposed DPP-SD scheme provides a significant reduction in computational complexity compared with the conventional SD algorithm, despite achieving near-optimal performance.
\end{abstract}


\maketitle

\section{Introduction}

In multiple-input multiple-output (MIMO) systems, the sphere decoding (SD) algorithm is known as an efficient signal-detection scheme, which performs close to the maximum-likelihood detection (MLD) receiver \cite{Hassibi}. 
Recently, to satisfy the increasing demand of ultra-high data rates in mobile communication systems, large MIMO systems, in which a large number of antennas are employed at a base station for data transmission and reception, have been of great research interest \cite{Elghariani}. To maximize the achievable data rates in large MIMO systems, the base station needs to receive as many symbols as possible simultaneously from multiple terminals, which leads to enhanced multiplexing gains. In this circumstance, a near-optimal receiver like SD plays an important role in approaching the channel capacity. However, the complexity of SD significantly increases with the number of antennas \cite{Seethaler}, which makes it difficult to apply to large MIMO systems.

Recently, deep-learning (DL) techniques have been applied in various fields, exhibiting eminent performance. Motivated by the performance of DL technologies in other fields, there have been attempts to apply DL to MIMO detection \cite{Mohammadkarimi}--\cite{He}. In particular, the DL-based sphere decoding (DL-SD) algorithm is derived to choose the optimal hypersphere radius \cite{Mohammadkarimi}. In addition, a deep network architecture, called DetNet, is proposed to estimate the solution of MIMO detection \cite{Samuel}. Furthermore, the sparsely connected neural network (ScNet) is developed to simplify the structure of DetNet for massive MIMO systems \cite{Gao}. The application of a deep neural network to reduce the computational complexity of the conventional belief propagation detector is proposed in \cite{Tan}, and the orthogonal approximate message-passing network (OAMP-Net) architecture is proposed to improve the performance of the iterative detection algorithm with trainable variables \cite{He}.

In this paper, a novel learning-aided SD algorithm is proposed. The main idea of the proposed algorithm is to predict the minimum path metric among ``deep'' paths of each sub-tree in a large tree structure by using a neural network (NN). In large MIMO systems, the required information to estimate the path metrics can have large dimension, which can significantly increase the complexity of the NN. To resolve this problem, the size of the input vector to the NN is optimized based on the property of large MIMO channels. The predicted minimum path metrics of sub-trees, which are generated by the designed NN architecture, are then used to determine the search order of SD. Furthermore, they are also used for early termination and optimization of the initial radius of SD, which can potentially reduce the overall complexity. 

The remaining part of this paper is organized as follows. In Section II, the system model and conventional Schnorr--Euchner SD (SE-SD) algorithm are explained, whereas Section III describes the proposed learning-aided SD algorithm, which employs the NN. Section IV presents the simulation results to compare the bit-error rate (BER) performance and computational complexity, and is followed by the conclusion in Section V.

\textit{Notations}: Scalars, vectors, and matrices are denoted by lowercase, bold-face lowercase, and bold-face uppercase letters, respectively. The $(i,j)$th element of a matrix $\textbf{A}$ is denoted by $a_{i,j}$, whereas the $i$th element of a vector $\textbf{a}$ is denoted by $a_{i}$. $(\cdot)^T$ and $(\cdot)^H$ represent the transpose and conjugate transpose of a matrix, respectively, whereas $\textbf{I}$ and $\textbf{0}$ indicate an identity matrix and all-zero matrix of appropriate size, respectively. $\Re$($\cdot$) and $\Im$($\cdot$) denote the real and imaginary parts of a complex matrix, respectively.

\section{System model}
We consider a MIMO system with ${N_t}$ transmit antennas and ${N_r}$ receive antennas. The received signal vector can be expressed as \\
\begin{equation} 
\mathbf {y = Hx + v},
\end{equation}      
where ${\mathbf y}$ represents an ${ N_r \times 1}$ complex received signal vector, ${\mathbf H}$ is an ${N_r \times N_t}$ complex channel matrix, ${\mathbf v}$ is an ${N_r \times 1}$ complex additive white Gaussian noise vector with zero mean and covariance matrix ${\mathbf \sigma_{v}^{2}\mathbf I}$, and ${\mathbf x}$ is an ${N_t \times 1}$ complex transmitted signal vector drawn from a QAM constellation ${\mathbb{S}}$. 
\par The optimal ML detector searches for the lattice point ${\mathbf {\hat{x}}_{ML}}$ that has the smallest Euclidean distance to the received signal vector ${\mathbf{y}}$ over the entire space ${\mathbb{S}}^{N_t}$ of the transmitted signal vector ${\mathbf x}$, yielding
\begin{equation} 
{\mathbf{\hat{x}}_{ML}=\arg\min\nolimits_{\mathbf {x\in \mathbb{S}}^{N_t}} \|\mathbf{ y- Hx}\|^{2}}.
\end{equation} 

In ML detection, all candidate vectors in $\mathbb{S}^{N_t}$ need to be examined, which requires high computational complexity, especially when $N_t$ is large. To resolve this problem, the SD algorithm can be employed.
To achieve near-ML performance with lower complexity, the SD algorithm limits the search space of the tree search. Specifically, for a search radius $d$, the SD solution can be expressed as

\begin{equation}\label{sd}
\mathbf{\hat{x}}_{SD}=\arg\min\nolimits_{ \big\{ \mathbf x\in \mathbb{S}^{N_t}\, \big | \,  \|\mathbf{y-Hx}\|^{2}\leq{d}^2 \big\} } \|\mathbf{y-Hx}\|^{2}.
\end{equation}
To improve the efficiency of the tree search procedure in SD, the QR decomposition (QRD) of the channel matrix ${\mathbf H}$ is performed, which yields \begin{equation}
\mathbf H = \mathbf Q \bigg[  {\begin{array}{cc}
   \mathbf R \\
   \mathbf 0 \\
  \end{array} } \bigg],    
\end{equation}
where  ${\mathbf R}$ is an ${N_t \times N_t}$ upper triangular matrix, ${\mathbf Q}=[\mathbf Q_1~\mathbf Q_2]$ is an ${N_r \times N_r}$ unitary matrix, and  $\mathbf Q_1$ and $\mathbf Q_2$ consists of the first $ N_t$ columns and last $(N_r-N_t)$ columns of $\mathbf Q$, respectively. Then, the constraint on the search space in (\ref{sd}) can be rewritten as 
\begin{equation}\label{qrd}
\|\mathbf{z-Rx}\|^{2}\leq \tilde{d}^2,
\end{equation} 
where we have ${\mathbf z=\mathbf Q_1^{H}\mathbf  y}$ and $\tilde{d}^2 = d^2-\|\mathbf {Q}_2^H\mathbf y\|^2$. Based on (\ref{qrd}), the SD solution can be reformulated as
\begin{equation}
\mathbf{\hat{x}}_{SD}=\arg\min\nolimits_{ \big\{ \mathbf x\in \mathbb{S}^{N_t} \,  \big | \,\|\mathbf{z-Rx}\|^{2}\leq \tilde{d}^2 \big\} } \|\mathbf{z-Rx}\|^{2}.
\end{equation}
\par The SE-SD scheme, known as an improved SD search strategy, determines search order for nodes at each layer based on the branch metrics \cite{Agrell}. For a candidate symbol vector $ {\mathbf x} = [{x}_1,~ {x}_2, ... ,~ {x}_{N_t}]^T$, the branch metric at layer \textit{l} is written as 
\begin{equation}\label{branch}
\textit{B}^{(l)}={\left({z}_l - \displaystyle\sum_{k=l}^{N_t} r_{l,k}{x}_k \right)}^2.
\end{equation}
In the SE-SD scheme, the candidate symbols are examined in ascending order of their branch metrics, which can be achieved at layer $l$ by following a zigzag search order of candidate symbols, starting from an initial point:
\begin{equation}
\bar{x}_l = \Bigg{\lfloor} \frac{1}{r_{l,l}} \displaystyle \left( z_l - \sum_{k=l+1}^{N_t} r_{l,k} x_k \right) \Bigg{\rceil},
\end{equation}
where $\lfloor \cdot \rceil$ rounds to the nearest point of its argument in the constellation $\mathbb{S}$. 

As aforementioned, at each layer of SE-SD, the search order is determined by the branch metric in (\ref{branch}), which only considers the metric at the corresponding layer. However, if we can use the full-path metric, i.e., $\|\mathbf{z-Rx}\|^{2}$, which accumulates the branch metrics from the root note to the leaf node, for ordering, the search can be more efficient. Furthermore, this full path metric can also be exploited for early termination and the optimization of the initial radius. This motivates us to develop a novel NN-based SD scheme, which utilizes the predicted path metric for the operation of SD. 

\section{Deep path prediction for sphere decoding}

In this section, the proposed DPP-SD is presented. As discussed in the last paragraph of Section II, knowledge of the full path metric can improve the efficiency of search ordering, which can potentially reduce the computational complexity. Hence, we consider the prediction of the path metric based on the NN. However, the prediction of the full path metric for every candidate vector requires the same complexity order as the ML detection, which implies that it is not an efficient strategy to exploit the NN to reduce the complexity of SD. Instead, in the proposed DPP-SD scheme, the NN is designed to predict the minimum path metric of the sub-tree rooted by each node at layer ${N_t}$. In other words, before beginning the tree search, the DPP-SD scheme predicts the minimum path metric of the ``deep path'' ranging from each child node of the root to a leaf node in each sub-tree, and we use it for sub-tree ordering, early termination, and radius determination.

Fig. 1 illustrates a tree structure for SD when quadrature phase shift keying (QPSK) modulation is employed, where we have $|\mathbb{S}|=4$ sub-trees, each of which are rooted by one of $|\mathbb{S}|$ nodes at layer ${N_t}$. In the next subsection, we will describe the NN that predicts the minimum path metrics of the sub-trees.

\begin{figure}[th] 
	\begin{center}
		\includegraphics[width=0.9\linewidth]{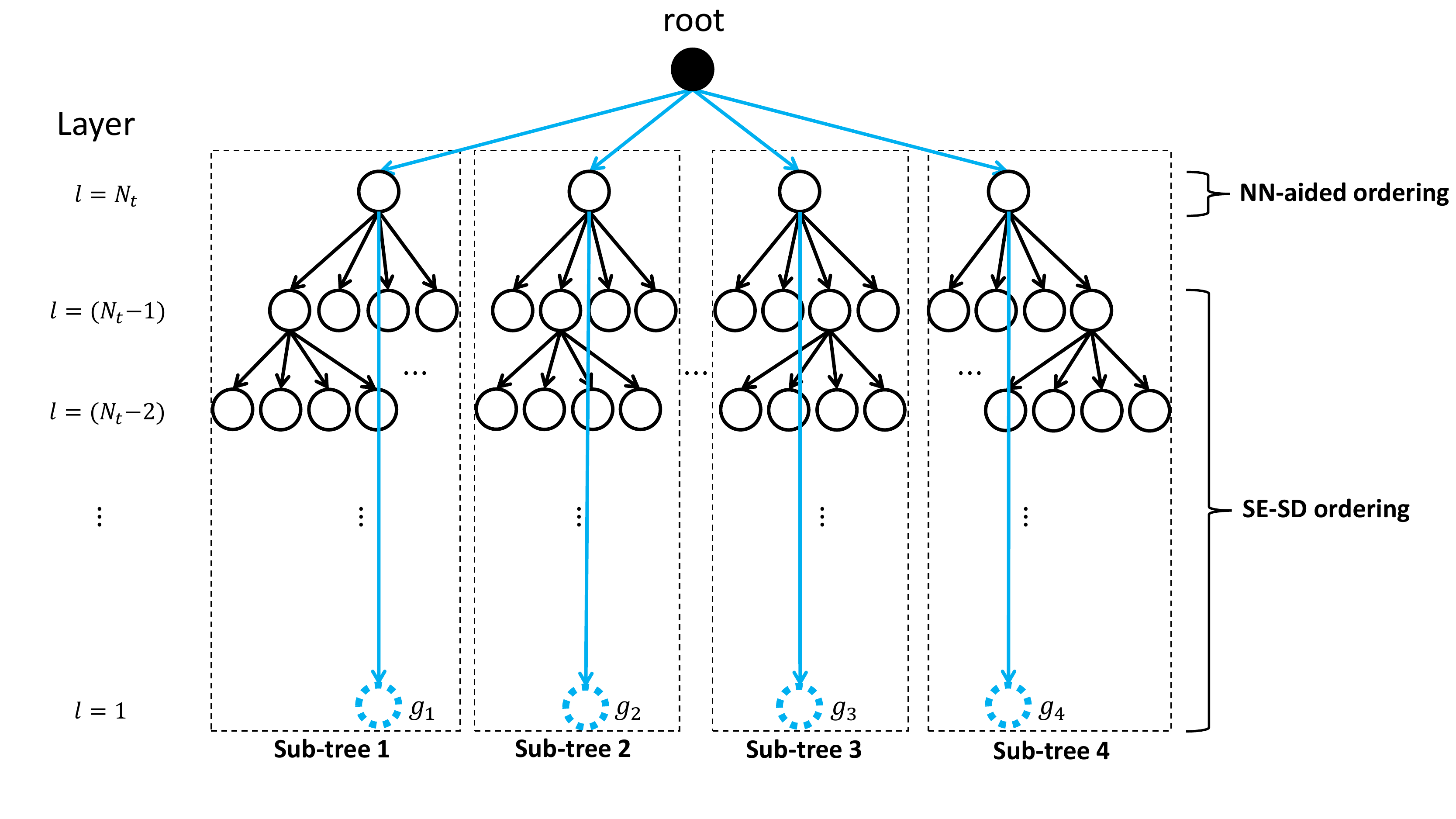}
	\end{center}
	\caption{Tree structure for $N_t\times N_r$ MIMO systems with QPSK.}
	\label{fig1:long}
	\label{fig1:onecol}
\end{figure}

\subsection{Design of the NN for path metric prediction}

Let $\mathbf{g} = [g_1,\,g_2,\,\cdots,\,g_{\vert\mathbb{S}\vert}]^T$ be the target vector of the proposed NN, where ${g}_q^2$ means the minimum path metric of the sub-tree rooted by the $q$th node at layer ${N_t}$, which can be formulated as
 \begin{equation}\label{target}
{g}_q^2 = ({z}_{N_t}-{r}_{N_t,N_t}{s}_q^{(N_t)})^{2}+\min_{\mathbf x_{1:N_t-1} \in \mathbb{S}^{N_t-1}}{\left(\displaystyle\sum_{k=1}^{N_t-1}\textit{B}^{(k)}\right)}. 
\end{equation}
Here, $\{ {s}_1^{(N_t)}$,\, ${s}_2^{(N_t)}$,\, $\cdots$ ,\, ${s}_{|\mathbb{S}|}^{(N_t)} \}$ are the candidate symbols at layer $N_t$, and $\mathbf x_{1:N_t-1}$ represents the vector consisting of the first $N_t-1$ symbols in $\mathbf x$, i.e., $\mathbf x_{1:N_t-1}=[{x}_1,~ {x}_2, ... ,~ {x}_{N_t-1}]^T$.

Because the path metric depends on the received signal vector and the channel matrix, the input to the NN for path metric prediction should include information of the received signals and channel coefficients. However, if all components of the received signals and channel state information, i.e., $\{\mathbf{y},~ \mathbf{H}\}$ or $\{\mathbf{z},~ \mathbf{R}\}$, are employed for the input, the complexity of the NN can be significantly large in large MIMO systems. Furthermore, a large number of input elements can require a large training set and lower the prediction accuracy.
\par To reduce the number of input elements, we rewrite the path metric as
\begin{eqnarray}\label{metric} 
{\|\mathbf{z}-\mathbf{R}\mathbf{x}\|}^2 &=& (\mathbf{z}-\mathbf{R}\mathbf{x})^{{H}}(\mathbf{z}-\mathbf{R}\mathbf{x}) \nonumber \\
&=& \mathbf{z}^{{H}}\mathbf{z}-\mathbf{x}^{{H}}\mathbf{R}^{{H}}\mathbf{z}-(\mathbf{R}^{{H}}\mathbf{z})^{{H}}\mathbf x \nonumber \\ && +\,\mathbf{x}^{{H}}\mathbf{R}^{{H}}\mathbf{R}\mathbf{x} ,
\end{eqnarray}
which implies that the path metric can be computed based on knowledge of $\mathbf z^{{H}}\mathbf z$ , $\mathbf R^{{H}}\mathbf z$, and $\mathbf{R}^H\mathbf{R}$. It also means that they can be used as the inputs to the NN to predict the path metrics, instead of the received signals and channel matrix. However, $\mathbf{R}^H\mathbf{R}$ contains many more elements than  $\mathbf z^{{H}}\mathbf z$ and  $\mathbf R^{{H}}\mathbf z$, whereas it can be approximated as a diagonal matrix $\mathbf{R}^H\mathbf{R}\approx N_r \sigma_h^2 \mathbf{I}$ in large MIMO systems due to the asymptotically favorable propagation and channel-hardening effect \cite{Ngo}, where $\sigma_h^2 = E(|h_{i,j}|^2)$. By assuming $\sigma_h^2=1$, $\mathbf{R}^H\mathbf{R}$ can be approximated as a fixed matrix, which is independent of either of the received signal or the channel information. In practical systems, $\sigma_h^2=1$ can be achieved by properly normalizing the received signal and channel matrix in (1). Hence, we exclude $\mathbf{R}^H\mathbf{R}$ from the set of inputs to the NN. We note that the distribution of the path metric depends on the noise variance ${\mathbf \sigma_{v}^{2}}$, which implies that the noise variance can help improve the accuracy of estimating the minimum path metric in the NN. Considering these aspects, we set the input vector in the form of
\begin{equation}
\mathbf{e} = [\mathbf z^{{H}}\mathbf z,\, \Re\{(\mathbf R^{{H}}\mathbf z)^T\},\, \Im\{(\mathbf R^{{H}}\mathbf z)^T\},\, \sigma_{v}^{2}]^T,
\end{equation}
where the size of $\mathbf{e}$ becomes $2N_t+2$. 

For the NN to predict the minimum path metrics, we employ a Gaussian radial basis function network (G-RBFN) \cite{Pislaru}, \cite{Chen} consisting of one hidden layer, as depicted in Fig. 2. The radial basis function is expressed as
\begin{equation}
\varphi(\gamma) = \exp\left(-\frac{(\gamma-\mu)^2}{\omega^2}\right),\quad \gamma\in\mathbb{R}.
\end{equation}
In the G-RBFN structure employed for the path metric prediction, the Gaussian function with center $\mu=0$ and width $\omega=1$ is used as the activation function in each node of the hidden layer. The number of nodes in the hidden layer is set to $2N_t+2\vert\mathbb{S}\vert$.

\begin{figure}[ht] 
	\begin{center}
		\includegraphics[width=0.8\linewidth]{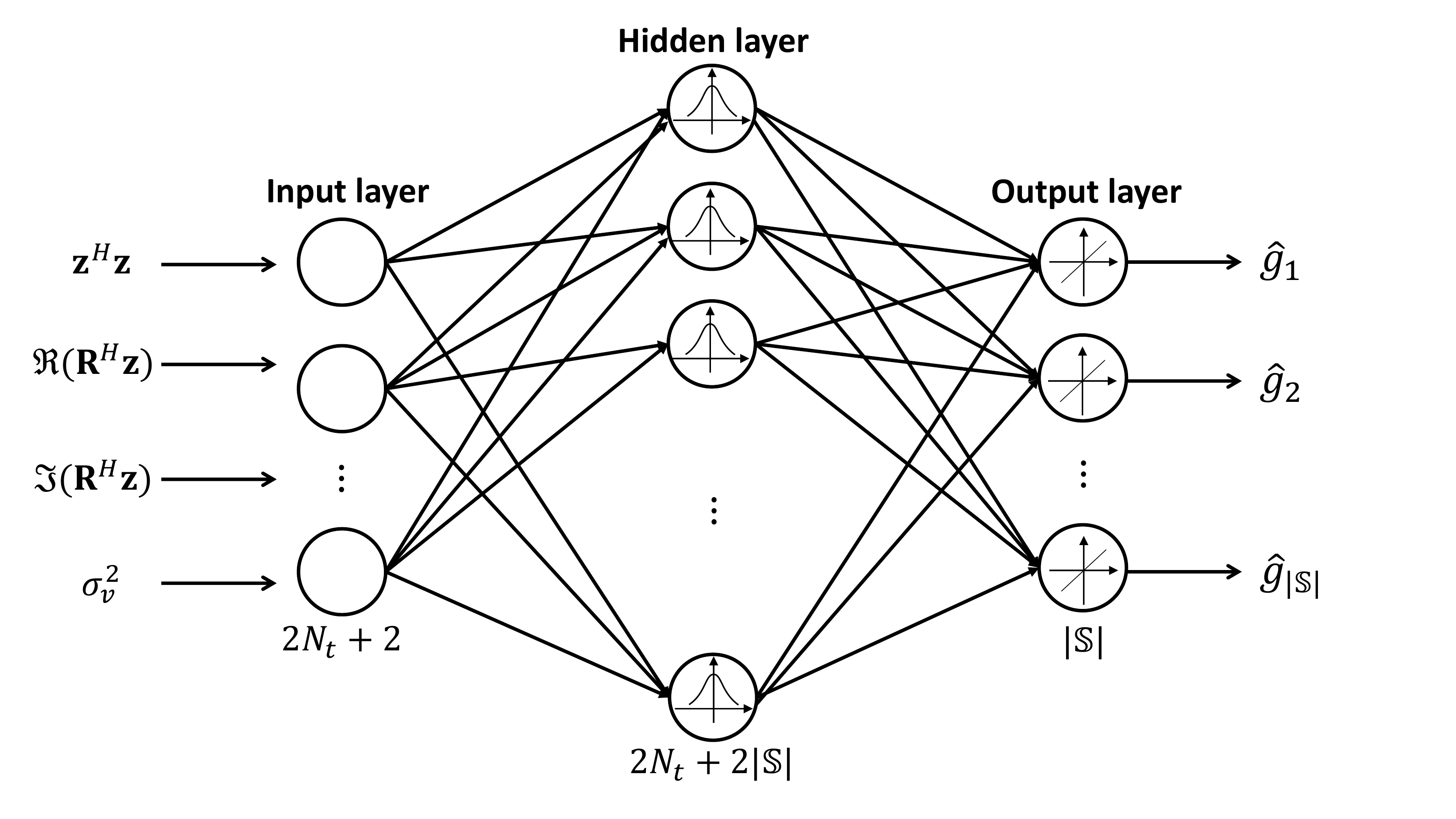}
	\end{center}
	\caption{The proposed NN architecture.\label{fig2}}
	\label{fig2:long}
	\label{fig2:onecol}
\end{figure}
 
 To optimize the parameter vector $\boldsymbol{\theta}$ of the NN, which consists of the weights and biases between input and hidden layers and between hidden and output layers, the mean squared error (MSE) loss function is used as follows:
\begin{equation}
\min_{\boldsymbol{\theta}}\textnormal{E}\left({\frac{1}{M}}\displaystyle\sum_{m=1}^{M}{\|\hat{\mathbf g}^{(m)}(\mathbf{e};\boldsymbol{\theta})-\mathbf{g}^{(m)}\|}^2\right),
\end{equation}
where $M$ indicates the number of training examples, whereas $\mathbf g^{(m)}$ and $\hat\mathbf{g}^{(m)}(\mathbf{e};\boldsymbol{\theta})=[\hat{g}^{(m)}_1,\,\hat{g}^{(m)}_2,\,\cdots,\,\hat{g}^{(m)}_{\vert\mathbb{S}\vert}]^T$ are the desired target vector and the output vector for the $m$th example. For the optimization algorithm in training, the scaled conjugate gradient (SCG) method \cite{Mller} is employed and the learning rate is set to 0.0001. 

\subsection{NN-aided optimization of the initial radius}
In the SD algorithm, the initial radius is typically determined based on the noise variance ${\mathbf \sigma_{v}^{2}}$ to meet the constraint on the probability of the true solution existing inside the sphere \cite{Hassibi}. Specifically, the conventional initial radius $f_{1-\epsilon}$ is chosen as

\begin{equation}
f_{1-\epsilon}=\sqrt{2N_r\Xi(1-\epsilon)\sigma_{v}^{2} },
\end{equation}
where $\Xi$ is the inverse incomplete gamma function, and $1-\epsilon$ is the probability of the true solution existing inside the sphere. In this work, we assume $1-\epsilon = 0.999$. If the initial radius is small, the complexity of SD is reduced; however, its BER performance can be degraded because the probability of the true solution being outside the sphere increases. In contrast, if the initial radius is large, the BER performance is improved, but the complexity can increase. Therefore, a better trade-off between the performance and complexity can be achieved if we reduce the initial radius while preserving the near-optimal BER performance.

We note that the NN presented in the prior subsection generates $\mathbf{\hat g}(\mathbf{e};\boldsymbol{\theta}) = [\hat g_1,\,\hat g_2,\,\cdots,\,\hat g_{\vert\mathbb{S}\vert}]^T$, which are the predicted smallest path metrics of all the sub-trees originating from layer $N_t$. Hence, we can consider $\tilde{g}_{1} = \min\{\hat g_1,\,\hat g_2,\,\cdots,\,\hat g_{\vert\mathbb{S} \vert }\}$ as the estimate of the smallest path metric over all the possible paths in the tree. Inspired by this, in the proposed DPP-SD scheme, we set the initial radius $\hat{f}$ to
\begin{equation}\label{radius}
\hat{f}=\min(\lambda_1 \tilde{g}_{1},\,f_{0.999}),
\end{equation}
where $\lambda_1$ is a design parameter. In the ideal case, where $\tilde{g}_{1}$ is accurately estimated and we have $\tilde{g}_1\leq f_{0.999}$, the initial radius $\hat{f}$ with $\lambda_1 = 1$ guarantees that there is at least a single lattice point inside the sphere, which implies that the proposed DPP-SD achieves the optimal ML performance. However, in practice, $\tilde{g}_1$ contains a prediction error, and hence $\lambda_1 >1$ is empirically chosen to provide near-optimal performance. From (\ref{radius}), we have $\hat{f} \leq f_{0.999}$, and hence we can expect that by setting the initial radius to $\hat{f}$ instead of $f_{0.999}$, the computational complexity of tree search can be reduced, as will be numerically verified in Section V.

\subsection{NN-aided ordering}

The output of the proposed NN can also be exploited for ordering in tree search. Specifically, the proposed NN-aided ordering scheme rearranges the predicted minimum path metrics in the output vector $\mathbf{\hat g}(\mathbf{e};\boldsymbol{\theta})$ in ascending order to generate $\tilde\mathbf{g}=[\tilde{g}_1,\,\tilde{g}_2,\,\cdots,\,\tilde{g}_{\vert\mathbb{S}\vert}]^T$, where we have $\tilde{g}_{1}\leq\tilde{g}_{2}\leq\cdots\leq\tilde{g}_{| \mathbb{S}|}$. The smaller predicted minimum path metric of a sub-tree implies that it is more likely that this sub-tree contains the final solution of SD. Therefore, it can be computationally efficient to search over the sub-trees with smaller predicted path metrics first. In the depth-first search of the proposed DPP-SD, we visit the nodes at layer $N_t$ in the order of $ \{\tilde{g}_1,\,\tilde{g}_2,\,\cdots,\,\tilde{g}_{\vert\mathbb{S}\vert}\}$. In other words, the sub-tree corresponding to $\tilde{g}_1$ is first visited by a depth-first search, and then the sub-tree corresponding to $\tilde{g}_2$ is searched. In this manner, the search is continued until any termination condition for SD is satisfied.

This NN-aided ordering strategy for layer $N_t$ can be extended to the remaining layers, which can potentially improve the search efficiency. However, it requires additional outputs of the NN. We note that the number of outputs of the NN increases exponentially with the number of layers employing the NN-aided ordering scheme, which leads to the significantly enhanced overall complexity of DPP-SD. For example, to apply the NN-aided ordering to layer $N_t -1 $, the NN should be designed to predict the minimum path metrics of $|\mathbb{S}|^2$ sub-trees rooted by the nodes at  layer $N_t -1 $. Considering this aspect, we only apply the NN-aided ordering scheme to layer $N_t$, whereas the ordering scheme of SE-SD is employed for the remaining layers, as illustrated in Fig. 1.

\subsection{NN-aided early termination }

To further reduce the complexity, early termination can be performed based on the predicted minimum path metrics. When a solution is found through the search for a sub-tree rooted by the $p$th node of the first layer, the radius $\hat{f}$ is updated to the distance of the found solution. After the search in the $p$th sub-tree is completed, the radius, which is generated by the current best solution, is compared to the predicted minimum path metric of the next sub-tree. In particular, we check the termination condition
\begin{equation}\label{et}
\lambda_{2} \hat{f}<\tilde{g}_{p+1},
\end{equation}
where $\lambda_2$ is a design parameter. If the condition (\ref{et}) is satisfied, the most recently found solution is adopted as the final solution and the algorithm is terminated, which results in complexity reduction of the SD. If there is no prediction error in $\tilde g_{q+1}$, we can choose $\lambda_{2} = 1$ in (\ref{et}) without any performance loss of the SD receiver. However, considering the potential prediction error in $\tilde{g}_{p+1}$, $\lambda_{2}$ should be set to be sufficiently large to guarantee the near-optimal performance of the SD.

The proposed DPP-SD scheme is summarized Algorithm 1. In step 1, we obtain the predicted path metrics $\mathbf{\hat g}(\mathbf{e};\boldsymbol{\theta})$ by using the trained NN. In steps 2 and 3, the sub-trees rooted by a node at layer $N_t$ are rearranged in ascending order of the elements in $\mathbf{\hat g}(\mathbf{e};\boldsymbol{\theta})$. Then, the NN-aided initial radius is determined based on the estimated smallest path metric $\tilde{g}_1$ in step 4. In steps 5--16, the tree search procedure is performed for each sub-tree. Specifically, in steps 6--10, the optimal solution in the $p$th sub-tree is determined. In steps 12--14, it is determined whether early termination is performed or not. 

\begin{algorithm}
\caption{The proposed DPP-SD algorithm}
\begin{algorithmic}[1]
\REQUIRE $\mathbf {z}=\mathbf Q_1^H{\mathbf y},~\mathbf R,~\mathbf{e},~\boldsymbol{\theta},~\lambda_1,~\lambda_2,~f_{0.999}$
\ENSURE $\hat{\mathbf x}$
\STATE Get $\mathbf{\hat g}(\mathbf{e};\boldsymbol{\theta})=[{\hat g}_1,\,{\hat g}_2,\,\cdots,\,{\hat g}_{\vert\mathbb{S}\vert}]^T$ by using the trained NN.
\STATE Sort $\mathbf{\hat g}(\mathbf{e};\boldsymbol{\theta})$ to generate $\tilde\mathbf{g}=[\tilde{g}_1,\,\tilde{g}_2,\,\cdots,\,\tilde{g}_{\vert\mathbb{S}\vert}]^T$.
\STATE Rearrange the sub-trees according to $\tilde\mathbf{g}$.
\STATE Set the initial radius: $\hat{f}=\min(\lambda_1\tilde{g}_{1},f_{0.999})$.
\FOR{$p=1 \,\, \TO \,\, \vert\mathbb{S}\vert$} \STATE {Perform the depth-first search in the $p$th sub-tree.} 
\IF {A new solution is found.}
\STATE Set the new solution to $\hat{\mathbf{x}}$.
\STATE Update the radius: $\hat{f}=\|\mathbf{z}-\mathbf{R} \hat{\mathbf{x}}\|$.
\STATE Go to Step 6 to continue the search for the $p$th sub-tree.
\ELSE \IF {$\lambda_{2}\hat{f}<\tilde{g}_{p+1}$}
\STATE Break. (Early Termination) 
\ENDIF
\ENDIF
\ENDFOR
\end{algorithmic}
\end{algorithm}

\section{Simulation results}
 	
In this section, the simulation results are presented to evaluate the performance and complexity of the proposed DPP-SD scheme. For simulations, we consider 16$\times$16 and 24$\times$24 MIMO systems with QPSK modulation. For each MIMO system, 100,000 randomly generated data samples are used to train the designed NN. To generate the random data samples, the channel coefficients are set to independent and identically distributed (i.i.d.) complex Gaussian random variables with zero mean and unit variance, whereas the signal-to-noise ratio (SNR) is set to a uniform random variable in the range of $[4,~14]$ dB. We define the SNR as $E_s N_t/\sigma_v^2$, where $E_s$ is the average symbol energy.

\begin{table}[th]
	\caption{Optimized values of $\lambda_1$ and $\lambda_2$.}
	\centering
	\def\sym#1{\ifmmode^{#1}\else\(^{#1}\)\fi}
	\scalebox{1.2}{%
	\begin{tabular}{l*{5}{c}}
		\toprule
		\multicolumn{1}{c}{SNR [dB]} &  \multicolumn{2}{c}{16${\times}$16 MIMO, QPSK}  & \multicolumn{2}{c}{24${\times}$24 MIMO, QPSK} \\
		\cmidrule(lr){2-3}\cmidrule(lr){4-5}
		& \multicolumn{1}{c}{$\qquad\lambda_1$} & \multicolumn{1}{c}{$\lambda_2$} & 
		\multicolumn{1}{c}{$\qquad\lambda_1$} & \multicolumn{1}{c}{$\lambda_2$}  \\
		\midrule
		\quad\,\,5 &\qquad1.2    &  1.3         &\qquad1.2  & 1.1       \\
		\addlinespace
		\quad\,\,7 &\qquad1.3    &  1.4         &\qquad1.3  & 1.2       \\
		\addlinespace
		\quad\,\,9 &\qquad1.4    &  1.5         &\qquad1.4  & 1.4      \\
		\addlinespace
		\quad11 &\qquad1.5   &  1.6         &\qquad1.4  & 1.5    \\
		\addlinespace
		\quad13 &\qquad1.6   &  1.7         &\qquad1.7  & 1.6   \\
		\bottomrule
	\end{tabular}%
	}
\end{table}

\begin{figure}[th] 
	\begin{center}
		\includegraphics[width=0.8\linewidth]{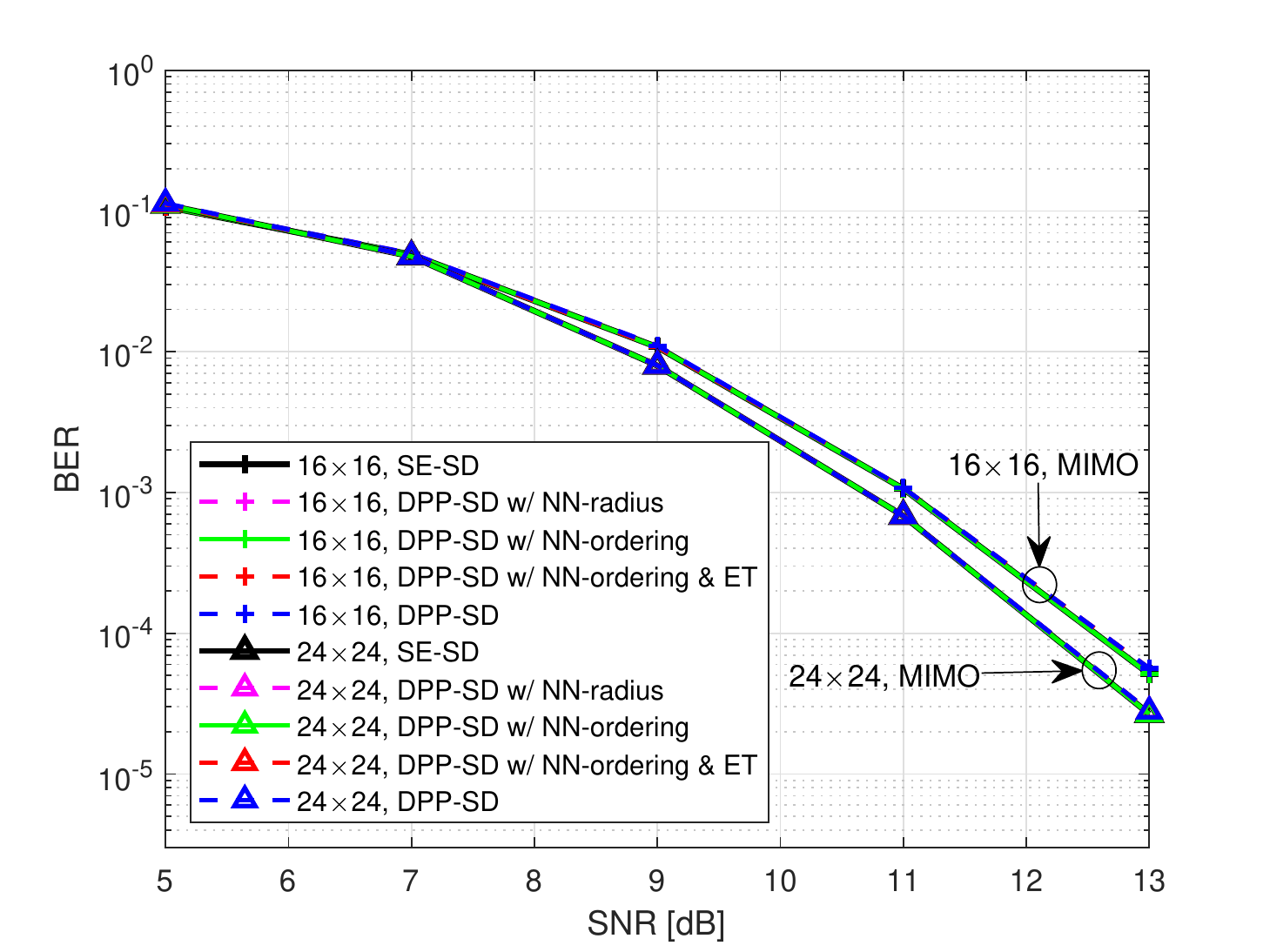}
	\end{center}
	\caption{BER performance comparison between DPP-SD and SE-SD for 16$\times$16 and 24$\times$24 MIMO systems with QPSK.\label{fig3}}
	\label{fig3:long}
	\label{fig3:onecol}
\end{figure}

\begin{figure}[th] 
	\begin{center}
		\includegraphics[width=0.8\linewidth]{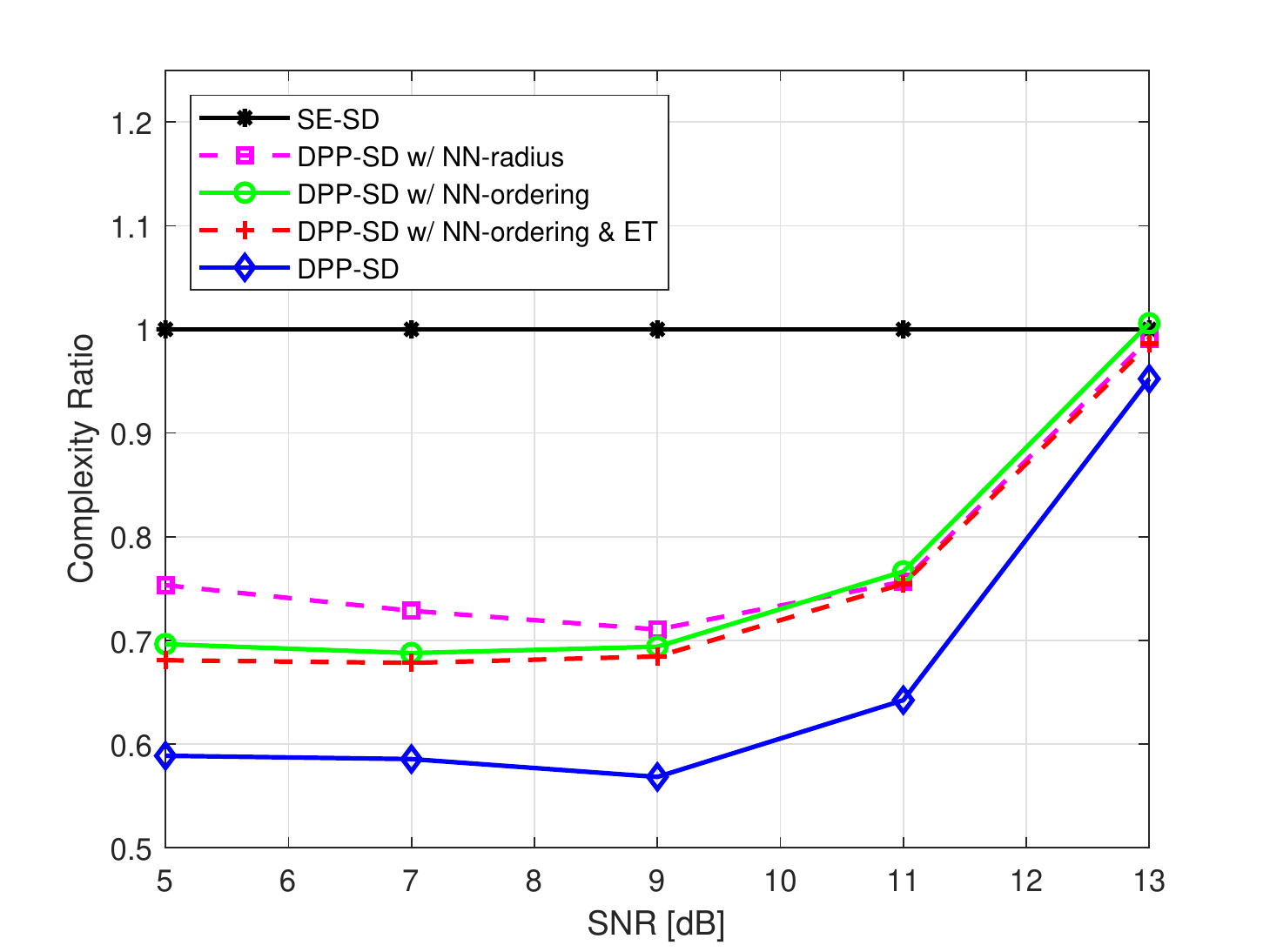}
	\end{center}
	\caption{Complexity comparison between the DPP-SD and SE-SD for a 16$\times$16 MIMO system with QPSK.\label{fig4}}
	\label{fig4:long}
	\label{fig4:onecol}
\end{figure}

\begin{figure}[th] 
	\begin{center}
		\includegraphics[width=0.8\linewidth]{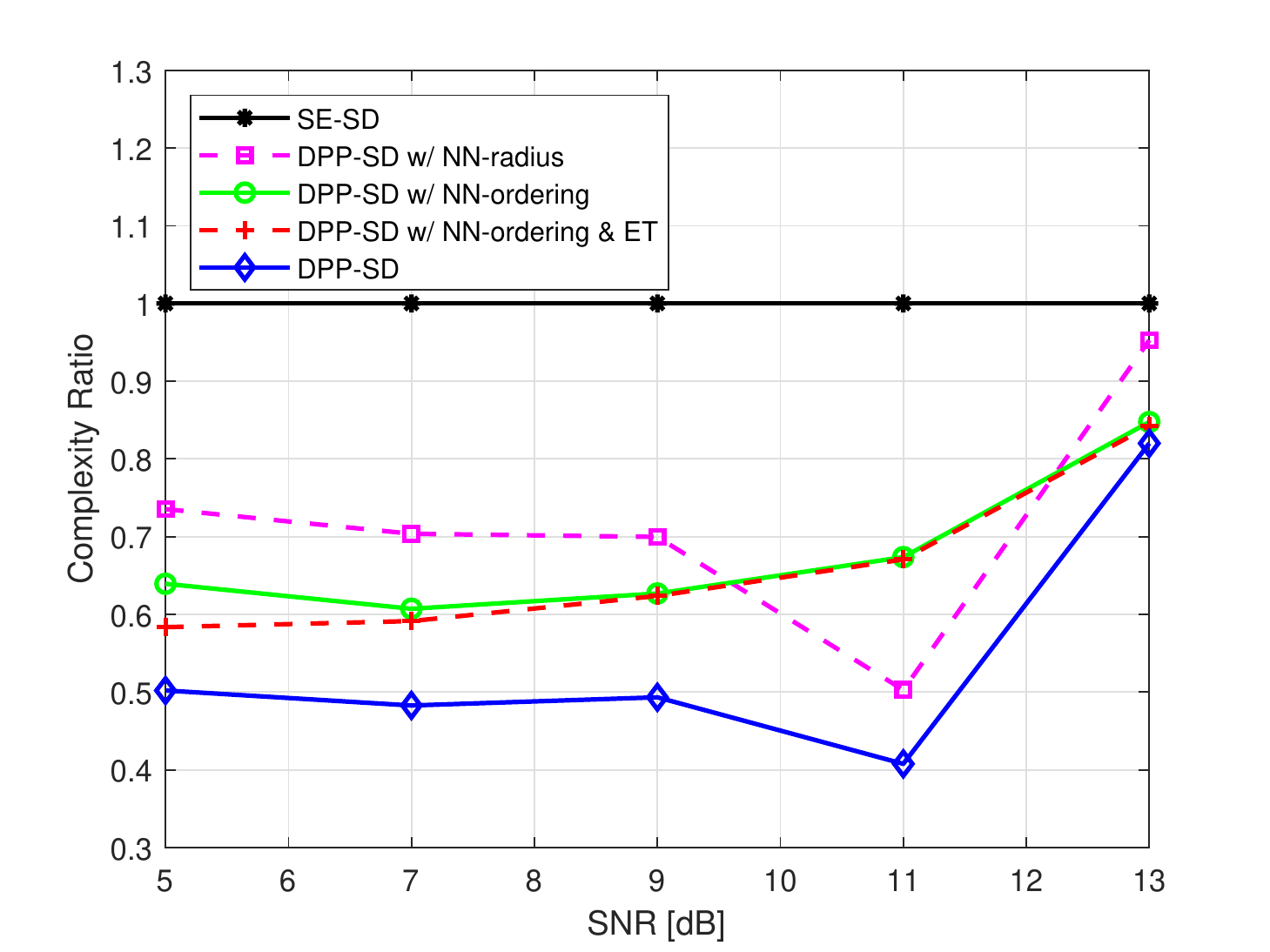}
	\end{center}
	\caption{Complexity comparison between the DPP-SD and SE-SD for a 24$\times$24 MIMO system with QPSK.\label{fig5}}
	\label{fig5:long}
	\label{fig5:onecol}
\end{figure}

The design parameters $\lambda_1$ and $\lambda_2$ of the proposed DPP-SD algorithm are optimized by simulations. As $\lambda_1$ and $\lambda_2$ increase, the performance improves; however, the complexity also increases. Therefore, the values of $\lambda_1$ and $\lambda_2$ are optimized such that the complexity of the DPP-SD is minimized while its performance remains nearly the same as that of the SE-SD. The optimized values of $\lambda_1$ and $\lambda_2$ for each MIMO configuration and SNR are shown in Table 1.

As presented in Section III, the DPP-SD scheme is composed of three sub-schemes: NN-aided initial radius, NN-aided sub-tree ordering, and early termination. To separately test these sub-schemes, we evaluate 1) DPP-SD with only the NN-aided initial radius, 2) DPP-SD with only NN-aided sub-tree ordering, and 3) DPP-SD with only NN-aided sub-tree ordering and early termination, which are referred to as ``DPP-SD w/ NN-radius,'' ``DPP-SD w/ NN-ordering,'' and ``DPP-SD w/ NN-ordering \& ET'' in the figures demonstrating the simulation results, respectively. We note that early termination can be employed only when it is incorporated with sub-tree ordering. For the conventional SD scheme, for comparison, the SE-SD algorithm is considered, whereas the DL-SD in \cite{Mohammadkarimi} is excluded from comparison. In the DL-SD algorithm, the original channel matrix and received signals are used for the input to the NN, which can have large dimension in large MIMO systems. For example, in a 24$\times$24 MIMO system, the input contains more than 1,000 elements, which requires significantly high complexity of the NN as well as a large training set. We note that in a 24$\times$24 MIMO system, the input vector of the proposed DPP-SD scheme only contains 50 elements.
In the considered algorithms, if no solution is found through the search process, the final solution is chosen to be a zero-forcing (ZF) solution, which is given by $\hat{\mathbf x}_{ZF}=(\mathbf H^H\mathbf H)^{-1}\mathbf H^H\mathbf{y}$.
  
Fig. 3 shows the BER performance comparison of the proposed DPP-SD and conventional SE-SD in the assumed MIMO configurations. In Fig. 3, it is observed that the proposed DPP-SD schemes achieve almost the same performance as the SE-SD.

\par In Figs. 4 and 5, we show the computational complexity ratio of the proposed DPP-SD with respect to the that of conventional SE-SD. To evaluate the computational complexity, the average number of complex multiplications and additions are counted. In Figs. 4 and 5, it is observed that the proposed DPP-SD requires significantly lower complexity than SE-SD when SNR $\leq$ 11 dB. In particular, the complexity-reduction ratio of the DPP-SD with all three sub-schemes is 36.7--43.2\% and 50--59.2\% at SNR $\leq$ 11 dB for 16$\times$16 MIMO and 24$\times$24 MIMO systems, respectively.

\section{Conclusion}
In this paper, we have presented the DPP-SD scheme, a novel learning-aided SD algorithm for large MIMO systems. To solve the high-complexity problem of the conventional SD algorithm in large MIMO systems, we design an NN to perform the minimum metric prediction among the paths in the sub-trees. To optimize the complexity of the NN, we reduce the size of the input vector based on the property of large MIMO channels. The DPP-SD algorithm exploits the output of the NN, i.e., the predicted minimum path metrics, to optimize the initial radius, sub-tree ordering, and early termination. The simulation results show that the proposed DPP-SD algorithm performs close to the conventional SE-SD scheme while requiring up to approximately 60\% lower complexity.

\ifCLASSOPTIONcaptionsoff
  \newpage
\fi


\end{document}